\documentclass[aps,prl,twocolumn,showpacs,preprintnumbers,amsmath,amssymb,superscriptaddress]{revtex4}%

\usepackage{graphicx}
\usepackage{dcolumn}
\usepackage{bm}
\usepackage{color}

\begin{document}

\preprint{preprint(\today)}

\title{Negative Oxygen Isotope Effect on the Static Spin Stripe Order in \\
La$_{2-x}$Ba$_{x}$CuO$_{4}$ ($x$ = 1/8)}

\author{Z.~Guguchia}
\email{zurabgug@physik.uzh.ch} \affiliation{Physik-Institut der
Universit\"{a}t Z\"{u}rich, Winterthurerstrasse 190, CH-8057
Z\"{u}rich, Switzerland}
\affiliation{Laboratory for Muon Spin Spectroscopy, Paul Scherrer Institute, CH-5232
Villigen PSI, Switzerland}

\author{R.~Khasanov}
\affiliation{Laboratory for Muon Spin Spectroscopy, Paul Scherrer Institute, CH-5232
Villigen PSI, Switzerland}

\author{M.~Bendele}
\affiliation{Physik-Institut der Universit\"{a}t Z\"{u}rich, Winterthurerstrasse 190, CH-8057
Z\"{u}rich, Switzerland}

\author{E.~Pomjakushina}
\affiliation{Laboratory for Developments and Methods, Paul Scherrer Institut, CH-5232 Villigen PSI, Switzerland}

\author{K.~Conder}
\affiliation{Laboratory for Developments and Methods, Paul Scherrer Institut, CH-5232 Villigen PSI, Switzerland}

\author{A.~Shengelaya}
\affiliation{Department of Physics, Tbilisi State University,
Chavchavadze 3, GE-0128 Tbilisi, Georgia}

\author{H.~Keller}
\affiliation{Physik-Institut der Universit\"{a}t Z\"{u}rich,
Winterthurerstrasse 190, CH-8057 Z\"{u}rich, Switzerland}

\begin{abstract}
Large negative oxygen-isotope ($^{16}$O/$^{18}$O) effects (OIE's) on the static spin-stripe ordering temperature $T_{\rm so}$ and 
the magnetic volume fraction $V_{\rm m}$ were observed in La$_{2-x}$Ba$_{x}$CuO$_{4}$ ($x$ = 1/8)
by means of muon spin rotation experiments.  
The corresponding OIE exponents were found to be $\alpha_{T_{\rm so}}$ = -0.57(6) and $\alpha_{V_{\rm m}}$ = -0.71(9), which
are sign reversed to $\alpha_{T_{\rm c}}$ = 0.46(6) measured for the superconducting transition temperature $T_{\rm c}$.
This indicates that the electron-lattice interaction is involved in the stripe formation and plays an important role
in the competition between bulk superconductivity and static stripe order in the cuprates.  

\end{abstract}


\maketitle

La$_{2-x}$Ba$_{x}$CuO$_{4}$ (LBCO) was the first cuprate system where high-$T_{\rm c}$ superconductivity was discovered \cite{Bednorz}.
This compound holds a unique position in the field since the bulk superconducting (SC) transition temperature $T_{\rm c}$ exhibits
a deep minimum at $x$ = 1/8 \cite{Moodenbaugh}, which is 
known as the 1/8 anomaly \cite{Kivelson,Vojta}. 
At this doping level neutron and X-ray diffraction experiments revealed two-dimensional static charge and spin (stripe) order  \cite{Tranquada1,Tranquada2,Abbamonte,Hucker}.
A central issue in cuprates is the microscopic origin of stripe formation and its relation to superconductivity. 
Given the fact that the amplitudes of the spin and charge orders as well as the ordering temperatures have maximum values at
$x$ = 1/8 \cite{Hucker}, where $T_{\rm c}$ is strongly suppressed, one might conclude that stripes and bulk (three-dimensional) superconductivity are
incompatible types of order. This conclusion is also supported by high-pressure muon spin rotation experiments (${\mu}$SR) in 
La$_{1.875}$Ba$_{0.125}$CuO$_{4}$ (LBCO-1/8) \cite{Guguchia}, demonstrating that static stripe order and bulk superconductivity  
occur in mutually exclusive spatial regions.
On the other hand, recent investigations of the relation between superconductivity and stripe order show that the situation 
is more complex, indicating quasi-two-dimensional superconductivity in LBCO-1/8, coexisting with static stripe order, 
but with frustrated phase order between the layers \cite{Tranquadareview,Tranquada2008,Li,Valla,Shen}. The frustrated Josephson coupling was explained in terms of
sinusoidally modulated [pair-density-wave (PDW)] SC order as proposed in Ref.~\cite{Berg1}. 
However, at present it is unclear to what extent PDW order is a common feature of cuprate systems 
where stripe order occurs. While the relevance of stripe correlations for high-temperature superconductivity remains a subject of controversy, the collected 
experimental data indicate that the tendency toward uni-directional stripe-like ordering
is common to cuprates \cite{Kivelson,Vojta,Kohsaka}. Exploring the mechanism of stripe formation will
help to clarify its role for the occurence of high-temperature superconductivity in the cuprates. 
The stripe phase may be caused by a purely electronic and/or electron-lattice interaction.
There is increasing experimental evidence for a strong electron-lattice interaction to be essential 
in the cuprates (see, e.g.,\cite{Mullerisotope,Kellerisotope,Kaimer}). 
However, it is not clear whether this interaction is involved in the formation of the stripe phase.

\begin{figure}[b!]
\centering
\includegraphics[width=1.0\linewidth]{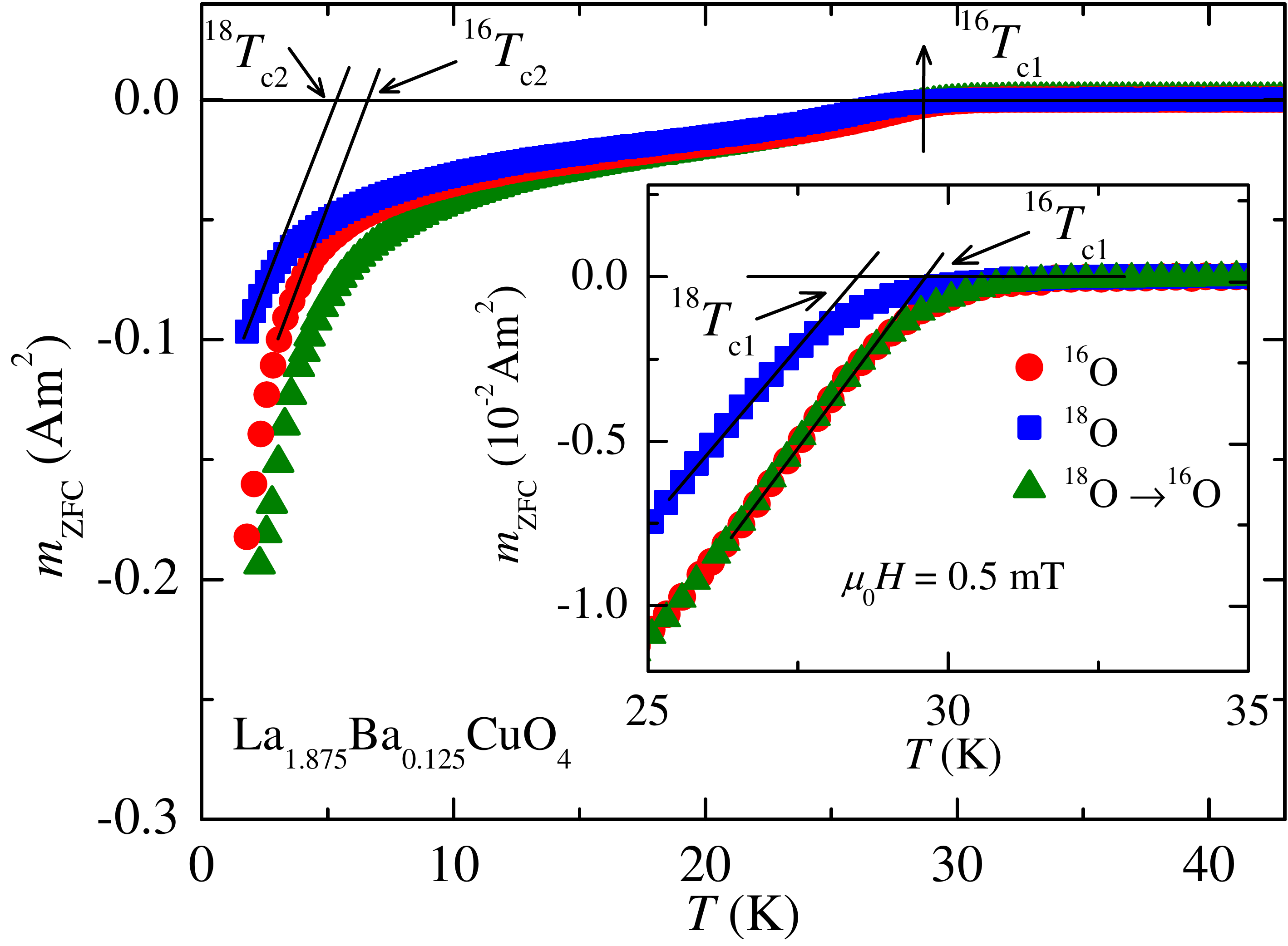}
\vspace{-0.6cm}
\caption{ (Color online) Temperature dependence of the diamagnetic moment $m_{\rm ZFC}$
for the $^{16}$O, $^{18}$O, and back-exchanged ($^{18}$O ${\rightarrow}$ $^{16}$O) samples of LBCO-1/8.
The arrows denote the superconducting transition 
temperatures $T_{\rm c1}$ and $T_{\rm c2}$ (see text for an explanation). The inset shows the SC transition near $T_{\rm c1}$.}
\label{fig1}
\end{figure}

Isotope effect experiments played  a crucial role for understanding superconductivity, since for conventional
superconductors they clearly demonstrated that the electron-phonon interaction is responsible for the electron pairing  \cite{Bardeen,Reynolds}.
In the cuprate high-temperature superconductors (HTS's) unconventional oxygen isotope ($^{16}$O/$^{18}$O) effects (OIE's) 
on various quantities were observed, such as the superconducting transition temperature $T_{\rm c}$,
the SC energy gap ${\Delta}$(0), the magnetic penetration depth ${\lambda}$(0), the N\'{e}el temperature
$T_{\rm N}$, the spin glass transition temperature $T_{\rm g}$, and the
pseudogap onset temperature $T^{*}$ \cite{Mullerisotope,Kellerisotope,shengelaya,khasanov2,Lanzara,Rubio}.
So far, a large OIE on $T_{\rm c}$ was observed in La$_{1.6-x}$Nd$_{0.4}$Sr$_{x}$CuO$_{4}$ \cite{Wang} and  
La$_{1.8-x}$Eu$_{0.2}$Sr$_{x}$CuO$_{4}$ \cite{Sury} showing stripe order at $x$ = 1/8  \cite{Crawford,Buchner}.
However, no OIE investigation on the charge and spin order in the stripe phase of cuprates has been reported.

In this letter we present OIE investigations of   
the static spin-stripe order in LBCO-1/8 by means of ${\mu}$SR experiments.
Substantial OIE's were found on magnetic quantities characterizing the static spin-stripe phase,
demonstrating that the electron-lattice interaction is essential in the stripe formation mechanism of cuprates.
In addition, we also studied the OIE on $T_{\rm c}$ in  LBCO-1/8 by magnetization measurements.
Remarkably, it was found that the OIE's have opposite signs for the magnetic and superconducting states 
in the stripe phase of LBCO-1/8. These findings reveal that lattice vibrations play an important role in the competition  
between superconductivity and static spin-stripe order in LBCO-1/8.

\begin{figure}[t!]
\includegraphics[width=1.0\linewidth]{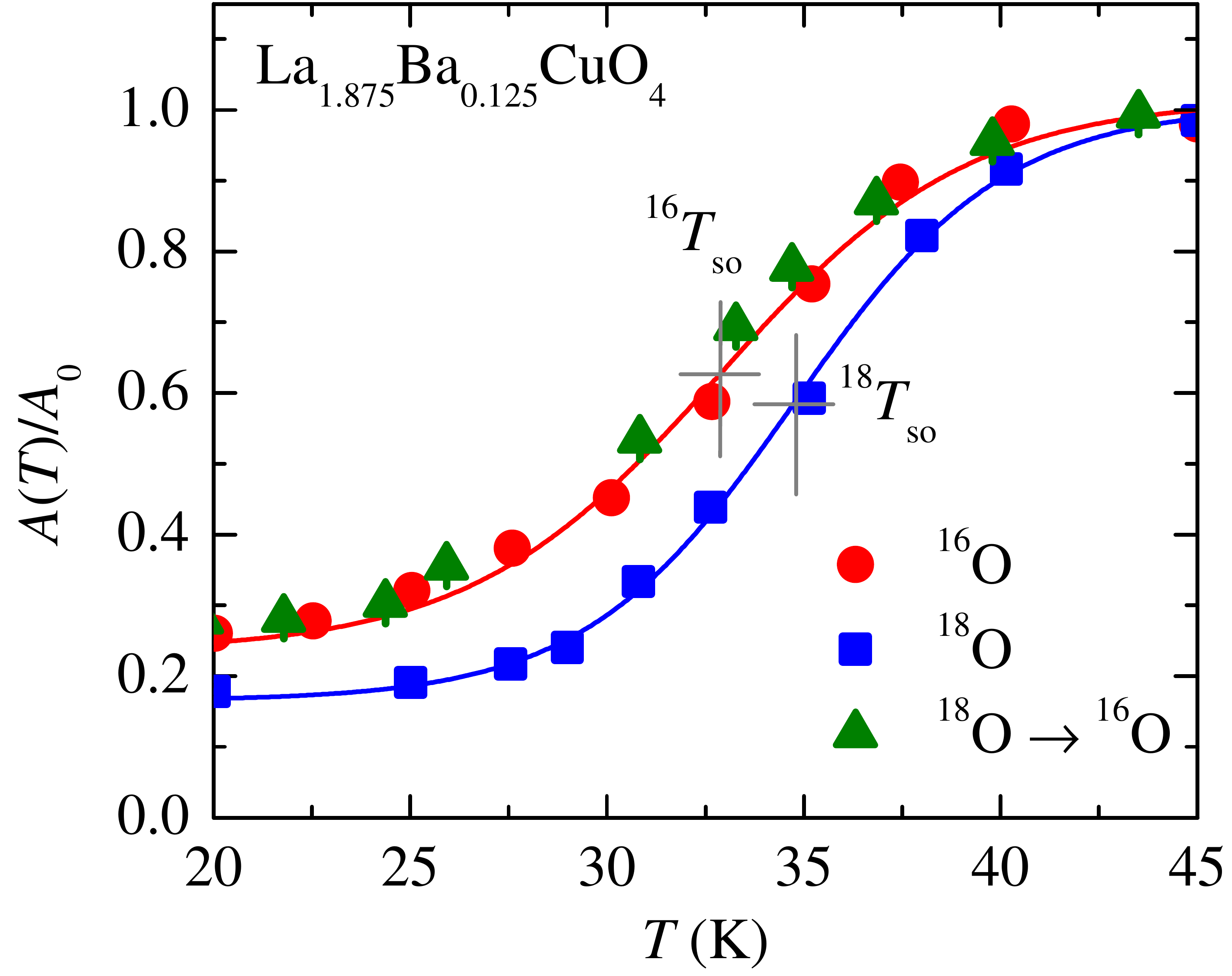}
\vspace{-0.7cm}
\caption{ (Color online) The normalized TF asymmetry $A$/$A_{\rm 0}$ plotted as 
a function of temperature for the $^{16}$O, $^{18}$O,
and  back-exchanged ($^{18}$O ${\rightarrow}$ $^{16}$O) samples  of LBCO-1/8. The crosses 
mark the spin-stripe order temperatures $^{16}$$T_{\rm so}$ and $^{18}$$T_{\rm so}$ 
for the $^{16}$O and $^{18}$O sample, respectively. The solid lines represent fits to the data by means of Eq.~(1).}\label{fig1}
\end{figure}

 A polycrystalline sample of La$_{2-x}$Ba$_{x}$CuO$_{4}$ with $x$ = 1/8 was prepared by 
the conventional solid-state reaction method using La$_{2}$O$_{3}$, BaCO$_{3}$, and CuO. The single-phase character of the sample was checked by powder x-ray diffraction.
All the measurements were performed on samples from the same batch. 
For the oxygen isotope exchange the sample was divided into two parts. To ensure that the substituted 
($^{18}$O) and not substituted ($^{16}$O) samples were subject of the same thermal history, 
both parts were annealed simultaneously in separate chambers (in $^{16}$O$_{2}$ and $^{18}$O$_{2}$ gas, respectively) under exactly the same conditions.
The oxygen isotope enrichment of the samples was determined in situ using mass spectrometry. The
$^{18}$O enriched samples contain ${\simeq}$ 82 ${\%}$ $^{18}$O and ${\simeq}$ 18 ${\%}$ $^{16}$O.

In a first step the OIE on the superconducting transition temperature $T_{\rm c}$ was determined by magnetization experiments performed with  a SQUID magnetometer ($Quantum$ $Design$ MPMS-XL) in a field of 0.5 mT.
The temperature dependence of the zero-field-cooled (ZFC) 
diamagnetic moment $m_{\rm ZFC}$ for the $^{16}$O, $^{18}$O, and back-exchanged ($^{18}$O ${\rightarrow}$ $^{16}$O) samples  of LBCO-1/8  
is shown in Fig.~1. The diamagnetic moment exhibits a two-step SC transition in all samples, similar to our previous work \cite{Guguchia}.
The first transition appears at $T_{\rm c1}$ ${\simeq}$ 30 K and the second transition at $T_{\rm c2}$ ${\simeq}$ 5 K 
with a larger diamagnetic response.
Detailed investigations performed on single crystalline samples of LBCO-1/8 provided an explanation for this two-step
SC transition \cite{Tranquada2008}. The authors interpreted the transition at $T_{\rm c1}$ as due to 
the development of 2D superconductivity in the CuO$_{2}$ planes, while the interlayer 
Josephson coupling is frustrated by static stripes. A transition to a 3D SC phase
takes place at much lower temperature $T_{\rm c2}$ ${\ll}$ $T_{\rm c1}$.  
The values of $T_{\rm c1}$ and $T_{\rm c2}$ were defined as the temperatures where the
linearly extrapolated magnetic moments intersect the zero line (see Fig.~1). 
Both $T_{\rm c1}$ and $T_{\rm c2}$ decrease by ${\simeq}$ 1.4 K and ${\simeq}$ 1.2 K, respectively, upon replacing 
$^{16}$O with $^{18}$O. 
To ensure that the observed changes of $T_{\rm c1}$ and $T_{\rm c2}$ are indeed due to isotope substitution, magnetization measurements were 
also carried out on a back-exchanged ($^{18}$O ${\rightarrow}$ $^{16}$O) sample (see Fig.~1). Note that the OIE on $T_{\rm c1}$
is very well reproducible (inset of Fig.~1). However, at low temperatures $m_{\rm ZFC}$($T$) for the back-exchanged sample does not 
follow the one for the $^{16}$O sample. This is due to the fact that the SC transition at $T_{\rm c2}$
is extremely sensitive to the thermal history (oxygen annealing time) of the samples,
which is about a factor of 2 longer for the back-exchanged sample.
Therefore, we only discuss the OIE on $T_{\rm c1}$ further. The following values for the OIE on $T_{\rm c1}$ 
were found: $^{16}$$T_{\rm c1}$ = 29.7(1) K, $^{18}$$T_{\rm c1}$ = 28.3(1) K, 
${\Delta}$$T_{\rm c1}$ = $^{18}$$T_{\rm c1}$ - $^{16}$$T_{\rm c1}$ = - 1.4(2) K, and for the OIE exponent
$\alpha_{T_{\rm c1}}$ = -dln$T_{\rm c1}$/dln$M_{\rm 0}$ = 0.46(6) ($M_{\rm 0}$ is the
oxygen isotope mass). Note that this value is comparable to that found for La$_{2-x}$Ba$_{x}$CuO$_{4}$ ($x$ = 0.10 - 0.15) \cite{Crawford2},
but is much smaller than $\alpha_{T_{\rm c}}$ ${\simeq}$ 1.89 for La$_{1.6-x}$Nd$_{0.4}$Sr$_{x}$CuO$_{4}$ ($x$ = 1/8) \cite{Wang}
and $\alpha_{T_{\rm c}}$ ${\simeq}$ 1.09 for La$_{1.8-x}$Eu$_{0.2}$Sr$_{x}$CuO$_{4}$ ($x$ = 0.16) \cite{Sury}. 
 
\begin{figure}[t!]
\includegraphics[width=1.0\linewidth]{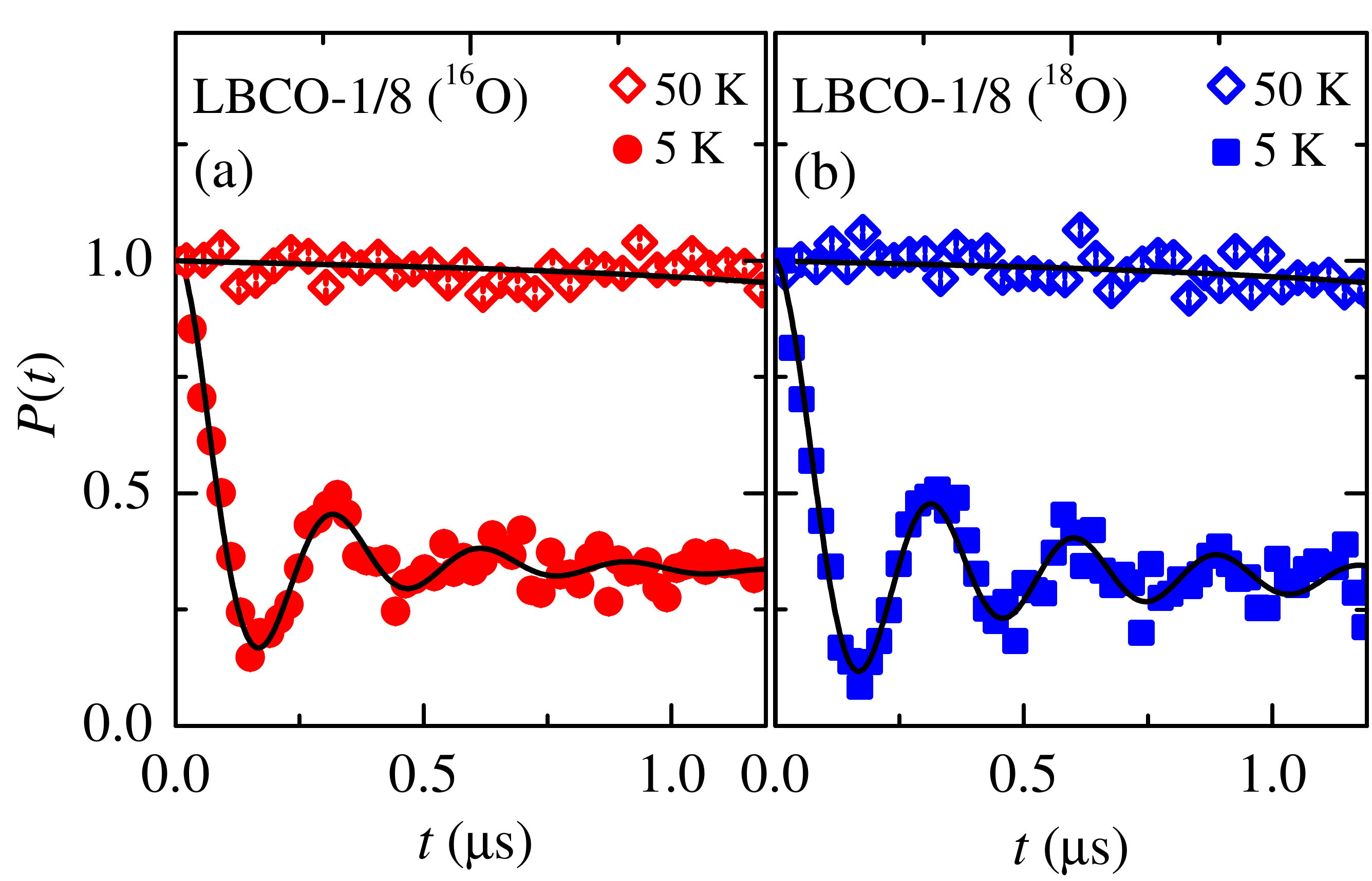}
\vspace{-0.7cm}
\caption{ (Color online)  ZF ${\mu}$SR signal $P(t)$ for the $^{16}$O (a) and $^{18}$O (b) samples  of LBCO-1/8 
recorded at $T$ = 5 K and $T$ = 50 K. 
The solid lines represent fits to the data by means of Eq.~(2).}
\label{fig1}
\end{figure}

Finally, the OIE on the static spin-stripe order in LBCO-1/8 was studied by means of zero-field (ZF) and 
transverse-field (TF) ${\mu}$SR experiments. 
In a ${\mu}$SR experiment positive muons implanted into a sample serve as an extremely
sensitive local probe to detect small internal magnetic fields and ordered magnetic volume fractions
in the bulk of magnetic materials. Note that the appearance of static magnetic order 
below ${\simeq}$ 30 K in LBCO-1/8 was originally observed by ${\mu}$SR \cite{Luke}.
The ${\mu}$SR experiments were carried out at the ${\pi}$M3 beam line at the Paul Scherrer Institute (Switzerland) using 
the general purpose instrument (GPS) with a standard veto setup providing a low-background ${\mu}$SR signal.
The ${\mu}$SR time spectra were analyzed using the free software package MUSRFIT \cite{Suter}.

Figure 2 shows the TF-${\mu}$SR asymmetry $A$ (normalized to its maximum value $A_{\rm 0}$), extracted
from the ${\mu}$SR spectra following the procedure given in Ref.~\cite{MBendele}, as a function of temperature 
for the $^{16}$O, $^{18}$O, and back-exchanged ($^{18}$O ${\rightarrow}$ $^{16}$O) samples of LBCO-1/8 in an applied field of ${\mu}_{0}$$H$ = 3 mT.
Above 40 K, $A$ saturates at a maximum value for both $^{16}$O and $^{18}$O, indicating that
the whole sample is in the paramagnetic state, and all the muon spins precess in the applied magnetic field.
Below 40 K, $A$ decreases with decreasing temperature and reaches an almost constant value at low temperatures.
The reduction of $A$ signals the appearance of magnetic order in the spin-stripe phase,
where the muon spins experience a local magnetic field larger than the applied magnetic field.
As a result, the fraction of muons in the paramagnetic state decreases.
Note that  $A$($T$) for the $^{18}$O sample is systematically shifted towards 
higher temperatures as compared to one for the $^{16}$O sample, indicating that the static spin-stripe ordering
temperature $^{18}$$T_{\rm so}$ for $^{18}$O is higher than $^{16}$$T_{\rm so}$ for $^{16}$O.
The values of $^{16}$$T_{\rm so}$ and $^{18}$$T_{\rm so}$ were determined by using the phenomenological function \cite{khasanov2}:  
\begin{equation}
A(T)/A_{0} = a\Bigg[1-\frac{1}{{\exp}[(T-T_{so})/{\Delta}T_{so}]+1}\Bigg]+b,
\label{eq1}
\end{equation}
where ${\Delta}T_{\rm so}$ is the width of the transition, and $a$ and $b$ are empirical parameters. 
Analyzing the data in Fig.~2 with Eq.~(1) yields:
$^{16}$$T_{\rm so}$ = 32.9(3) K and  $^{18}$$T_{\rm so}$ = 34.8(2) K 
with a large negative OIE exponent $\alpha_{T_{\rm so}}$ = -0.56(9).
A back exchange experiment ($^{18}$O ${\rightarrow}$ $^{16}$O) was carried out
in order to exclude any doping differences in the oxygen-isotope exchanged samples.
As shown in Fig. 2 the oxygen  
back-exchanged sample of LBCO-1/8 exhibits within experimental error almost the same $A$($T$) as the 
$^{16}$O sample. This demonstrates that the observed negative OIE on $T_{\rm so}$ is intrinsic.
Note that $\alpha_{T_{\rm so}}$ = -0.56(6) and $\alpha_{T_{\rm c1}}$ = 0.46(6) have almost the same
magnitude, but \textit{sign reversed}. 
\begin{table*}[ht!]
\caption{The values of $T_{\rm so}$, ${\Delta}$$T_{\rm so}$ = $^{18}$$T_{\rm so}$ - $^{16}$$T_{\rm so}$, 
$V_{\rm m}$(0), and ${\Delta}$$V_{\rm m}$(0) = $^{18}$$V_{\rm m}$(0) - $^{16}$$V_{\rm m}$(0) of the $^{16}$O/$^{18}$O samples of LBCO-1/8 determined from various measured
${\mu}$SR parameters. The OIE exponents ${\alpha}$$_{T_{\rm so}}$ and ${\alpha}$$_{V_{\rm m}}$ are corrected for the incomplete $^{18}$O exchange of 82(5)${\%}$.}
\vspace{0.3cm}
\begin{tabular*}{1.0\linewidth}{@{\extracolsep{\fill}}cccccccccccccc}
\hline
\hline  
Parameter             & $^{16}$$T_{\rm so}$ & $^{18}$$T_{\rm so}$ & ${\Delta}$$T_{\rm so}$ & ${\alpha}$$_{T_{\rm so}}$ & $^{16}$$V_{\rm m}$(0) & $^{18}$$V_{\rm m}$(0) & ${\Delta}$$V_{\rm m}$(0)  & ${\alpha}$$_{V_{\rm m}}$ \\ \hline
$A$($T$)     & 32.9(3) & 34.8(2) & 1.9(4) & -0.56(9) & ... & ... & ... & ... & \\ 
$B_{\rm \mu}$($T$)      & 30.1(3) & 31.8(3) & 1.7(5) & -0.55(11) & ... & ... & ... & ... \\ 
$V_{\rm m}$($T$)        & 35.2(2) & 37.4(2) & 2.2(3) & -0.61(7) & 0.82(1) & 0.88(1) & 0.06(1) & -0.71(9) \\ \hline
\hline         
\end{tabular*}
\label{table1}
\end{table*}
  
In order to explore the OIE on the magnetic volume fraction $V_{\rm m}$ as well as on $T_{\rm so}$, 
ZF ${\mu}$SR experiments (no external magnetic field applied) were carried out. Figure~3 shows representative ZF ${\mu}$SR time spectra for 
the $^{16}$O and $^{18}$O samples of LBCO-1/8. Below $T$ ${\approx}$ 30 K damped oscillations due to the presence of a
local magnetic field at the muon site are observed, indicating long range static spin-stripe order \cite{Luke,Nachumi}.
The ${\mu}$SR signals in the whole temperature range were analyzed by
decomposing the signal into a magnetic and a nonmagnetic contribution \cite{Nachumi}: 
\begin{equation}
\begin{split}
P(t)=V_{m}\Bigg[{\frac{2}{3}e^{-\lambda_{T}t}J_0(\gamma_{\mu}B_{\mu}t)}+\frac{1}{3}e^{-\lambda_{L}t}\Bigg] \\ 
+(1-V_{m})e^{-\lambda_{nm}t}.
\label{eq1}
\end{split}
\end{equation}
Here, $P$(t) is the muon spin polarization function. 
$V_{\rm m}$ denotes the relative magnetic volume fraction, and $\gamma_{\mu}/(2{\pi}) \simeq 135.5$~MHz/T 
is the muon gyromagnetic ratio. 
$B_{\rm \mu}$ is the average internal magnetic field at the muon site. ${\lambda_T}$ and ${\lambda_L}$
are the depolarization rates representing the transversal and the longitudinal 
relaxing components related to the spin-stripe ordered regions of the sample, respectively.
$J_{0}$ is the zero$^{th}$-order Bessel function of the first kind.
This is characteristic for an incommensurate spin-density wave 
and has been observed in cuprates with static spin-stripe order \cite{Nachumi}.
${\lambda_{nm}}$ is the relaxation rate related to the nonmagnetic part of the sample, where spin-stripe order is absent.

\begin{figure}[t!]
\includegraphics[width=1.0\linewidth]{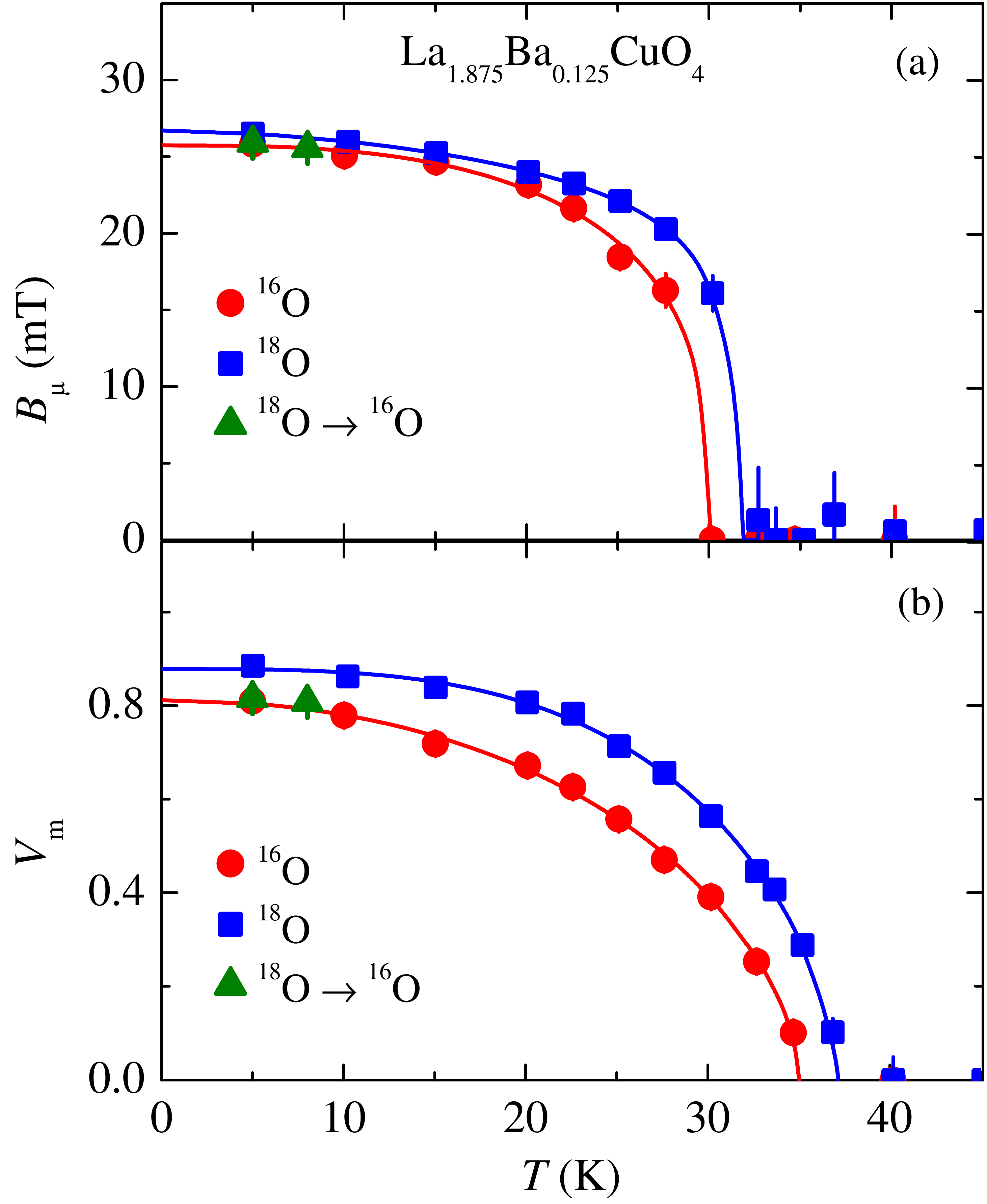}
\vspace{-0.8cm}
\caption{ (Color online) (a) Temperature dependence of the average internal 
magnetic field $B_{\rm \mu}$ at the muon site for  $^{16}$O, $^{18}$O,
and  back-exchanged ($^{18}$O ${\rightarrow}$ $^{16}$O) samples  of LBCO-1/8. 
The solid lines represent fits of  the data to the power law described in the text.   
(b) The temperature dependence of the magnetic 
volume fraction $V_{\rm m}$ for  $^{16}$O, $^{18}$O, and  back-exchanged ($^{18}$O ${\rightarrow}$ $^{16}$O) samples  of LBCO-1/8.
The solid lines are fits of the data to the same empirical power law as used for $B_{\rm \mu}$($T$) in (a).}  
\label{fig1}
\end{figure}

The temperature dependence of the average internal magnetic field $B_{\rm \mu}$ for the $^{16}$O, $^{18}$O, 
and back-exchanged samples of LBCO-1/8 is shown in Fig.~4a. It is evident that in the $^{18}$O sample $B_{\rm \mu}$ appears at a higher temperature than in the
$^{16}$O sample, showing that $^{18}$$T_{\rm so}$ is higher than $^{16}$$T_{\rm so}$.
The solid curves in Fig.~4a are fits of the data to the power law 
$B_{\rm \mu}$($T$) = $B_{\rm \mu}$(0)[1-($T/T_{\rm so}$)$^{\gamma}$]$^{\delta}$,
where $B_{\rm \mu}$(0) is the zero-temperature value of $B_{\rm \mu}$. ${\gamma}$ and ${\delta}$ are phenomenological exponents.
The analysis yields $^{16}$$T_{\rm so}$ = 30.1(3) K, $^{18}$$T_{\rm so}$ = 31.8(3)K, and
the OIE exponent of $T_{\rm so}$ obtained from $B_{\rm \mu}(T)$ is  $\alpha_{T_{\rm so}}$ = -0.55(11).

${\mu}$SR also allows to determine the magnetic volume fraction $V_{\rm m}$ in magnetically ordered materials. 
Figure~4b shows the temperature dependence of $V_{\rm m}$ for the $^{16}$O and $^{18}$O samples.
The solid lines in Fig.~4b are fits of the data to the same empirical power law as used for $B_{\rm \mu}(T)$ discussed above.
The OIE exponent of $T_{\rm so}$ obtained from $V_{\rm m}(T)$ is  $\alpha_{T_{\rm so}}$ = -0.61(7), 
in excellent agreement with $\alpha_{T_{\rm so}}$ = -0.55(11) and $\alpha_{T_{\rm so}}$ = -0.56(9) obtained from
the temperature dependence of the ${\mu}$SR parameters $B_{\rm \mu}$ and $A$, respectively.
This demonstrates that the two independent ${\mu}$SR experiments, TF and ZF ${\mu}$SR,   
give consistent results for $\alpha_{T_{\rm so}}$, although the values of $T_{\rm so}$ are systematically different (see Table~I) \cite{Comment}.
For further discussions we use the average value  $<$$\alpha_{T_{\rm so}}$$>$ = $\alpha_{T_{\rm so}}$ = -0.57(6) 
determined from the three measured values.
It is also clear from Fig.~4b that $V_{\rm m}$ in the $^{18}$O sample is significantly larger than in the $^{16}$O sample in the whole temperature
range, indicating a higher volume fraction of the static spin-stripe order phase in the $^{18}$O sample.
The zero-temperature values of the magnetic volume fraction were found to be $^{16}$$V_{\rm m}$(0) = 0.82(1) and $^{18}$$V_{\rm m}$(0) = 0.88(1),
yielding an OIE exponent of $\alpha_{V_{\rm m}}$ = -dln$V_{\rm m}$/dln$M_{0}$ = -0.71(9).
As shown in Fig.~4b the intrinsic OIE on $V_{\rm m}$(0) was confirmed by
back-exchange ($^{18}$O ${\rightarrow}$ $^{16}$O) experiments. 
The obtained results show that the quantities $T_{\rm so}$ and $V_{\rm m}$(0) characterizing the static spin-stripe state 
exhibit a large and negative OIE.
To our knowledge this is the first study reporting a substantial OIE on the static spin-stripe order state in a 1/8 doped cuprate.

 The values of $T_{\rm so}$ and $V_{\rm m}$(0) related to the static spin-stripe phase of $^{16}$O/$^{18}$O exchanged LBCO-1/8
obtained in this work as well as the corresponding OIE exponents are summarized in Table~I.
The average value of the static spin-stripe order temperature $T_{\rm so}$ ${\simeq}$ 33 K is
in agreement with the previous values $T_{\rm so}$ ${\simeq}$ 30 - 34 K obtained from ${\mu}$SR \cite{Luke,Nachumi} and comparable to the
value of the superconducting transition temperature $T_{\rm c1}$ ${\simeq}$ 30 K.
However, the value of $T_{\rm so}$ determined by ${\mu}$SR is smaller than $T_{\rm so}$ ${\simeq}$ 40 K determined by
neutron scattering \cite{Tranquada2008} due to the different time window of the two techniques.
One should point out that the values of $T_{\rm so}$ and $V_{\rm m}$(0) 
increase with increasing oxygen-isotope mass (Figs.~2 and 4), whereas
$T_{\rm c1}$ decreases (Fig.~1). This demonstrates a competition between bulk superconductivity and static spin-stripe order in LBCO-1/8, 
and that the electron-lattice coupling is involved in this competition.

  In conclusion, oxygen isotope effects on magnetic and superconducting quantities
related to the static stripe phase of LBCO-1/8 were investigated by means of ${\mu}$SR and magnetization experiments.
The static spin-stripe order temperature $T_{\rm so}$ and the magnetic volume fraction $V_{\rm m}$(0) exhibit
a large negative OIE which is novel and unexpected. 
This indicates that the electron-lattice interaction plays an essential role for the stripe formation in cuprate HTS's. 
Furthermore, the observed oxygen-isotope shifts of the superconducting transition temperature $T_{\rm c1}$ and the
spin-ordering temperature $T_{\rm so}$ have almost the same magnitude, but opposite signs.
This provides clear evidence that bulk superconductivity and static spin-order
are competitive phenomena in the stripe phase of LBCO-1/8, and that the electron-lattice interaction is a  
crucial factor controlling this competition. The present results may contribute to a better
understanding of the complex microscopic mechanism of stripe formation and of high-temperature superconductivity in the cuprates
in general.

We are grateful to A.~Bussmann-Holder for valuable discussions. 
The ${\mu}$SR experiments were performed at the Swiss Muon Source, Paul Scherrer Institute (PSI),
Villigen, Switzerland. This work was supported by the Swiss National Science Foundation, the NCCR MaNEP, 
the SCOPES grant No. IZ74Z0-137322, and the Georgian National Science Foundation grant RNSF/AR/10-16.


\begin{thebibliography}{22}

\bibitem{Bednorz} J.G. Bednorz and K.A. M\"{u}ller, Z. Phys. B \textbf{64}, 189 (1986).

\bibitem{Moodenbaugh} A.R. Moodenbaugh, Y. Xu, M. Suenaga, T.J. Folkerts, and R.N. Shelton,
Phys. Rev. B \textbf{38}, 4596 (1988).

\bibitem{Kivelson} S.A. Kivelson, I.P. Bindloss, E. Fradkin, V. Oganesyan, J.M. Tranquada, A. Kapitulnik, and C. Howald,
Rev. Mod. Phys. \textbf{75}, 1201 (2003).

\bibitem{Vojta} M. Vojta, Adv. Phys. \textbf{58}, 699 (2009).



\bibitem{Tranquada1} J.M. Tranquada, B.J. Sternlieb, J.D. Axe, Y. Nakamura, and S. Uchida, 
Nature (London) \textbf{375}, 561 (1995).

\bibitem{Tranquada2} J.M. Tranquada, J.D. Axe, N. Ichikawa, Y. Nakamura, S. Uchida, and B. Nachumi, 
Phys. Rev. B \textbf{54}, 7489 (1996).

\bibitem{Abbamonte} P. Abbamonte, A. Rusydi, S. Smadici, G.D. Gu, G.A. Sawatzky, and D.L. Feng,  Nat. Phys. \textbf{1}, 155 (2005).

\bibitem{Hucker} M. H\"{u}cker,  M. v. Zimmermann, M. Debessai, J.S. Schilling, J.M. Tranquada, and G.D. Gu,
Phys. Rev. Lett. \textbf{104}, 057004 (2010).

\bibitem{Guguchia} Z. Guguchia, A. Maisuradze, G. Ghambashidze, R. Khasanov, A. Shengelaya,
and H. Keller, New Journal of Physics \textbf{15}, 093005 (2013). 

\bibitem{Tranquadareview} J.M. Tranquada, AIP Conference Proceedings \textbf{1550}, 114 (2013).

\bibitem{Tranquada2008} J.M. Tranquada, G.D. Gu, M. H\"{u}cker, Q. Jie, H.-J. Kang, R. Klingeler, Q. Li, N. Tristan, 
J.S. Wen, G.Y. Xu, Z.J. Xu, J. Zhou, and M. v. Zimmermann, Phys. Rev. B \textbf{78}, 174529 (2008).

\bibitem{Li} Q. Li, M. H\"{u}cker, G.D. Gu, A.M. Tsvelik, and J.M. Tranquada, Phys. Rev. Lett. \textbf{99}, 067001 (2007).

\bibitem{Valla} T. Valla, A. V. Federov, J. Lee, J. C. Davis, and G. D. Gu, Science \textbf{314}, 1914 (2006).

\bibitem{Shen} R.-H. He, K. Tanaka, S.-K. Mo, T. Sasagawa, M. Fujita, T. Adachi, N. Mannella, K. Yamada,
Y. Koike, Z. Hussain, and Z.-X. Shen, Nat. Phys. \textbf{5}, 119–123 (2009).

\bibitem{Berg1} E. Berg, E. Fradkin, E.-A. Kim, S. A. Kivelson, V. Oganesyan, J. M. Tranquada, and S. C. Zhang,
Phys. Rev. Lett. \textbf{99}, 127003 (2007).

\bibitem{Kohsaka} Y. Kohsaka, C. Taylor, K. Fujita, A. Schmidt, C. Lupien, T. Hanaguri, M. Azuma,
M. Takano, H. Eisaki, H. Takagi, S. Uchida, and J. C. Davis, Science \textbf{315}, 1380 (2007).

\bibitem{Mullerisotope} K.A. M\"{u}ller, J.~Phys.~Condens.~Matter \textbf{19}, 251002 (2007).

\bibitem{Kellerisotope} H. Keller, A. Bussmann-Holder, and K.A. M\"{u}ller, Materials Today  \textbf{11}, 9  (2008);
H. Keller, and A. Bussmann-Holder, Advances in Condensed Matter Physics 
(Volume 2010, Article ID 393526, 17 pages, doi:10.1155/2010/393526).

\bibitem{Kaimer} M. Le Tacon, A. Bosak, S.M. Souliou, G. Dellea, T. Loew, R. Heid, K-P. Bohnen, G. Ghiringhelli,
M. Krisch, and B. Keimer, Nature Phys. \textbf{10}, 52-58 (2014). 

\bibitem{Reynolds} C.A. Reynolds, B. Serin, W.H. Wright, and L.B.  Nesbitt, Phys. Rev. \textbf{78}, 487 (1950);
E. Maxwell, Phys. Rev. \textbf{78}, 477 (1950).

\bibitem{Bardeen}  J. Bardeen, L.N. Cooper, and J.R. Schrieffer, Phys. Rev. \textbf{108}, 1175 (1957). 

\bibitem{shengelaya} A. Shengelaya, Guo-meng Zhao, C.M. Aegerter, K. Conder, I.M. Savi\'{c}, and H. Keller, Phys. Rev. Lett. \textbf{83}, 24  (1999).

\bibitem{khasanov2} R. Khasanov, A. Shengelaya, D. Di Castro, E. Morenzoni, A. Maisuradze, I.M. Savi\'{c}, K. Conder, 
E. Pomjakushina, A. Bussmann-Holder, and H. Keller, Phys. Rev. Lett. \textbf{101}, 077001  (2008).

\bibitem{Lanzara} A. Lanzara, Guo-meng Zhao, N.L. Saini, A. Bianconi, K. Conder,
H. Keller, and M\"{u}ller, J. Phys.: Condens Matter \textbf{11}, L541 (1999).

\bibitem{Rubio} D. Rubio Temprano, J.Mesot, S. Janssen, K. Conder, A. Furrer, H. Mutka, and K.A.M\"{u}ller, 
Phys.~Rev.~Lett. \textbf{84}, 1990 (2000).

\bibitem{Wang} G.Y. Wang, J.D. Zhang, R.L. Yang, and X.H. Chen, Phys. Rev. B \textbf{75}, 212503 (2007).

\bibitem{Sury} Suryadijaya, T. Sasagawa, and H. Takagi, Physica C \textbf{426}, 402 (2005).

\bibitem{Crawford} M.K. Crawford, R.L. Harlow, E.M. McCarron, W.E. Farneth, J.D. Axe, H. Chou, and Q. Huang, Phys. Rev. B \textbf{44}, 7749 (1991).

\bibitem{Buchner} B. B\"{u}chner, M. Breuer, A. Freimuth, and A.P. Kampf, Phys. Rev. Lett. \textbf{73}, 1841 (1994).

\bibitem{Crawford2} M.K. Crawford, M. Kunchur, W.E. Farneth, E.M. McCarron, and S.J. Poon, Physica C \textbf{162-164}, 755 (1989).

\bibitem{Luke} G.M. Luke, L.P. Le, B.J. Sternlieb, W.D. Wu, Y.J. Uemura, J.H. Brewer, T.M. Riseman, S. Ishibashi, and S. Uchida, 
Physica C \textbf{185-9}, 1175 (1991).

\bibitem{Suter} A. Suter and B.M. Wojek, $Physics~Procedia$ \textbf{30}, 69-73 (2012).

\bibitem{MBendele} M. Bendele, P. Babkevich, S. Katrych, S.~N. Gvasaliya, E.~Pomjakushina, K.~Conder, B. Roessli, A.~T. Boothroyd,
R. Khasanov, and H. Keller, Phys.~Rev.~B \textbf{82}, 212504 (2010).

\bibitem{Nachumi} B. Nachumi, Y. Fudamoto, A. Keren, K.M. Kojima, M. Larkin, G.M. Luke, J. Merrin, O. Tchernyshyov, Y.J. Uemura, 
N. Ichikawa, M. Goto, H. Takagi, S. Uchida, M.K. Crawford, E.M. McCarron, D.E. MacLaughlin, and R.H. Heffner,
Phys. Rev. B \textbf{58}, 8760-8772 (1998).

\bibitem{Comment} The reason for this might be that close to the static stripe phase transition magnetic fluctuations 
give rise to a depolarized ${\mu}$SR signal with a non-zero value for the magnetic fraction.
In the static spin-stripe ordered phase a well defined field is sensed by the muon spin.

        

\end{thebibliography}
\end{document}